\documentstyle[aps,prc,epsfig]{revtex}
\begin{document}
\draft

\wideabs{
\title{Period of the $\gamma$-ray staggering in the $^{150}$Gd
superdeformed region}
\author{D. Samsoen,$^1$ P. Quentin,$^1$ and I. N. Mikha\"{\i}lov$^{2,3}$}
\address{$^1$Centre d'\'{E}tudes Nucl\'{e}aires de Bordeaux Gradignan, 
CNRS-IN2P3 and Universit\'{e} Bordeaux-I, F-33175 Gradignan, France \\
$^2$Bogoliubov Laboratory of Theoretical Physics (Joint Institute of
Nuclear Research), 141980 Dubna (Moscow region), Russia \\ 
$^3$Centre de Spectrom\'{e}trie Nucl\'{e}aire et de Spectrom\'{e}trie de Masse
CNRS-IN2P3, 91405 Orsay, France}
\maketitle

\begin{abstract}
It has been previously proposed to explain $\gamma$-ray staggerings in the
deexcitation of some superdeformed bands in the $^{150}$Gd region in terms of
a coupling between global rotation and intrinsic vortical modes. The observed
4$\hbar$ period for the phenomenon is suggested from our microscopic Routhian
calculations using the Skyrme SkM$^*$ effective interaction.
\end{abstract}
}

\narrowtext

This brief paper completes the theoretical investigation of the coupling
between global rotation and intrinsic vortical modes proposed in
Ref. \cite{PRL:stagg} as a tentative explanation for the rather rare band
staggering observed in the decay of some superdeformed bands. Indeed while
such a phenomenon was first claimed in $^{149}$Gd \cite{Flibotte}, then in
$^{194}$Hg \cite{Cederwall}, and possibly in the $A\sim 130$ superdeformed
region \cite{Semple}, its existence has been confirmed and extended to a
couple of neighboring nuclei in the first case \cite{Haslip} and ruled out in
the second case \cite{Krucken}. Various theoretical explanations have been
proposed \cite{Hamamoto,Pavlichenko,Sun,Kota,Toki} besides the one which we
have discussed in Ref. \cite{PRL:stagg}. When making an attempt to describe
such data, one should address the three following questions: (i) What is the
mechanism at work? (ii) Why is this phenomenon so scarce and what makes it
appear where it is observed ($A\sim150$)? (iii) What is tuning the period
of the staggering?

While some answers have been provided in our previous papers
\cite{PRL:stagg,Nupha,Aharonov,Iouldash} to the two first questions, we aim
here at addressing the third one. In \cite{PRL:stagg}, a staggering in
transition energies within the yrast band was shown to appear in cases where
the relevant collective energy is quadratic in two quantized quantities. A
particular realization of the latter, well suited to the description of fastly
rotating superdeformed states, corresponds to the parallel coupling of global
rotation and intrinsic vortical modes in ellipsoidally deformed bodies, known
after Chandrasekhar \cite{Chandra} as \emph{S} ellipsoids. In this case, the
two commuting operators are the projections on the quantification axis of the
angular momentum operator and of the so-called Kelvin circulation operator
(see, e.g., Ref. \cite{Rosensteel}), hereafter called $I$ and $J$,
respectively. Whereas it is trivial to show that the Kelvin circulation
operator satisfies the usual commutation relations of an angular momentum, its
consideration as a quantity which is approximately a constant of the motion is
a basic assumption of our collective model. Its exact amount of violation
would deserve a specific microscopic study. A self-consistent description of
such a coupling can be made upon generalizing the Routhian approach
\cite{Routhian}, amounting thus to solve the following variational problem:
\begin{equation}
\delta\langle H -\Omega I - \omega J \rangle = 0,
\end{equation}
where $H$ is the microscopic Hamiltonian, $\Omega$ ($\omega$) an angular
velocity associated with the global rotation (intrinsic vortical)
mode. This approach was first investigated in Ref. \cite{Nupha} within a
simple oscillator mean field approximation. There, the assumption of an energy
which is quadratic in ($\Omega,\omega$) or equivalently in ($I,J$) has been
shown to be rather well satisfied. Recently, fully self-consistent solutions
of the above variational problem have been made possible \cite{Triaxial} upon
using the standard SkM$^*$ force \cite{Skmstar}. Details about such
calculations and some of their results will be discussed below.

In another paper \cite{Aharonov}, a physical analogy stemming from the
well-known similarity between the motion of a charge in a magnetic field and
of a mass in a rotating frame has been established. It relates this staggering
phenomenon with the observation of persistent currents in mesoscopic conductor
or semiconductor rings as a manifestation of an inherent Aharonov-Bohm
phase. Apart from its obvious physical appeal this consideration has set a
framework in which one is able to understand the scarcity of both
phenomena. It results from the necessity of securing a sufficiently low level
of a specific damping which happens to yield in the staggering case a condition
on the width of the superdeformed state in the relevant collective variable
(e.g., the usual axial quadrupole deformation $\beta$). It was deduced from
microscopic calculations of the associated mass parameters \cite{Iouldash},
using the usual D1S Gogny effective force \cite{D1S}, that such a condition of
existence was generally not met for Ce and Hg superdeformed states as well as
for such states in Gd isotopes but for $^{150}$Gd and possibly $^{148}$Gd.

As a consequence of all these studies the explanation for the staggering
phenomenon suggested in Ref. \cite{PRL:stagg} is of course neither exempted
from \textit{a priori} questions nor deemed as being the only possible one,
yet it is conforted as a rather likely candidate. However, one point remains to
be clarified concerning the $4\hbar$ period of the staggering. In
Refs. \cite{PRL:stagg,Nupha} this particular period was associated with a
ratio of $I$ and $J$ values close to 2 for the considered states. From either
semiclassical estimates of the relevant inertia parameters \cite{PRL:stagg} or
from actual microscopic calculations in the harmonic oscillator mean field
approximation \cite{Nupha} such a ratio is much more consistent with
hyperdeformation than with the actual superdeformation of the nuclear
states. It is on this point that the new self-consistent approach of
Ref. \cite{Triaxial} brings some interesting insight.

Here the generalized Routhian variational problem is solved within the
Hartree-Fock approximation. Numerical codes breaking the time-reversal and
axial symmetries as requested by the considered physical problem, determine
the single-particle wave functions either at the nodes of a spatial mesh
\cite{Bonche:87} or as resulting from an expansion on a suitably chosen
truncated basis. In the latter case, all approaches so far
\cite{Egido,Girod,Dudek}, to the best of our knowledge, have used a triaxial
basis. In our calculations an alternative method, shown to be less
time consuming in most cases, has been developed where the expansion is made
on an axial basis. Here the dependence of various fields and densities in
terms of the angular variable of the cylindrical coordinate system is handled
by convenient Fourier expansions.

\begin{figure}
\begin{center}
\epsfig{angle=-90,width=8cm,file=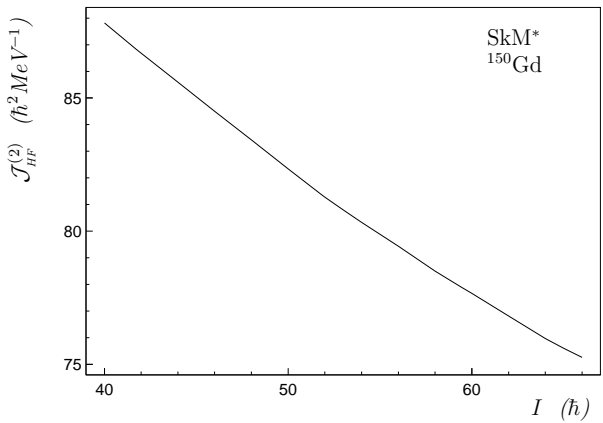}
\end{center}
\caption{Dynamical moment of inertia ${\cal J}^{(2)}$ as a function of the
angular momentum $I$.}
\label{fig:1}
\end{figure}

Self-consistent solutions of the usual Routhian problem (i.e. without
constraint on the Kelvin circulation, namely, for $\omega=0$) have been first
performed for the $^{150}$Gd nucleus. As usual to get a solution corresponding
to a given (quantized) value of $I$, one solves the problem iteratively so as
to determine the angular velocity constraint $\Omega$ able to reach the
requested angular momentum. For such solutions, one can estimate the dynamical
moment of inertia by numerical differentiation. As suggested in
Ref. \cite{Gall}, among the possible expressions for ${\cal J}^{(2)}$, we
choose for numerical stability reasons to evaluate
\begin{equation}
{\cal J}^{(2)} = \frac{\partial I}{\partial \Omega} .
\end{equation}
Such derivatives have been determined from a two-point formula involving thus
for each studied value of the angular momentum three different Hartree-Fock
calculations (differing typically around $I=50\hbar$ by $\Delta\hbar\Omega\sim
0.07$ keV). As discussed, e.g., in Ref. \cite{Nupha}, even in the absence of
any constraint on the intrinsic vortical currents, the Kelvin circulation
operator takes a finite value which is easily estimated from our
solutions. The vector components of this operator are defined (see, e.g.,
Ref. \cite{Rosensteel}) by its action on single-particle wave functions as
\begin{equation}
J_\alpha = \frac{\hbar}{i} \sum_{\beta,\gamma} \epsilon_{\alpha\beta\gamma}
\frac{c_\gamma}{c_\beta} x_\beta \frac{\partial}{\partial x_\gamma} ,
\end{equation}
where $\epsilon_{\alpha\beta\gamma}$ is the completely antisymmetrical 
third-rank tensor, $x_\alpha$ the $\alpha$ component of the particle position
vector, and $c_\alpha$ the corresponding length scale factor. These factors are
deduced from the values of the quadrupole tensor calculated from our
variational solutions upon making an ellipsoidal shape approximation.

Some results of our Routhian calculations ($\omega=0$) for a range of
$I$ values from $I=40\hbar$ to $I=66\hbar$, including states of relevance for
the observed superdeformation, are displayed in Fig. \ref{fig:1}. It shows the
variation of the moment of inertia ${\cal J}^{(2)}$ as a function of $I$. As a
matter of fact, it fits rather well, as it should, with the pure Hartree-Fock
part of the results obtained in Ref. \cite{Bonche:96} with the same
interaction. The calculated yrast value $J_{\mathrm{yrast}}(I)$ of the Kelvin
circulation as a function of $I$ falls very nicely on the straight line:
\begin{equation}
J_{\mathrm{yrast}}(I) \simeq 0.8 I + 1.0 \hbar .
\end{equation}
This is not at all surprising insofar as the quadratic approximation for the
collective energy discussed in Ref. \cite{Nupha} is valid over the whole
considered range of values of $I$. However, this relation makes it very clear
that around $I=50\hbar$, e.g., the ratio $I/J$ is indeed very far from the
value of 2 which would lead to the observed staggering period.

\begin{figure}
\begin{center}
\epsfig{angle=-90,width=8cm,file=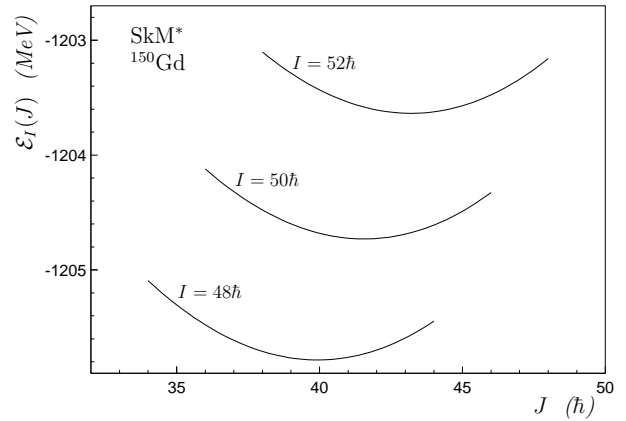}
\end{center}
\caption{Total energy ${\cal E}_I(J)$ as a function of the Kelvin circulation
$J$ for angular momentum values around $I=50\hbar$.}
\label{fig:3}
\end{figure}

As a result of the continuity equation, tangential intrinsic vortical
excitations yield phase-space modifications amounting only to a momentum
redistribution \cite{Nupha}. In that respect they are indeed quite similar to
pairing correlations. This has been, for instance, illustrated from another
point of view in Ref. \cite{Durand}. There, current patterns as functions of a
fixed pairing gap display indeed the same type of variation as classical
\emph{S}-ellipsoid velocity fields with respect to the angular velocity of the
intrinsic vortical modes. Therefore one may consider the latter modes as a
collective model translation of pairing correlations in a somewhat similar
fashion as small amplitude vibrational collective modes can model random phase
approximations (RPA) correlations.

In order to implement this type of excitation on top of Hartree-Fock
rotational solutions we have solved the generalized Routhian problem with two
constraints ($\Omega$ and $\omega$, both $\neq0$). Indeed we have made
calculations for fixed values of $I$ upon varying $J$. Clearly the single
constraint corresponds to the yrast state (up to the quantization of $J$ of
course). It is therefore no surprise to find it at the minimum of a somewhat
parabolic energy curve ${\cal E}_I(J)$ as demonstrated in Fig. \ref{fig:3}. 
Note in passing that the general pattern of such energy curves ${\cal E}_I(J)$ 
for various values of $I$ is consistent with a quadratic dependence of the 
total energy in $I$ and $J$ as assumed in Ref. \cite{PRL:stagg} and calculated
in the simple model case of Ref. \cite{Nupha}.

\begin{figure}
\begin{center}
\epsfig{angle=-90,width=8cm,file=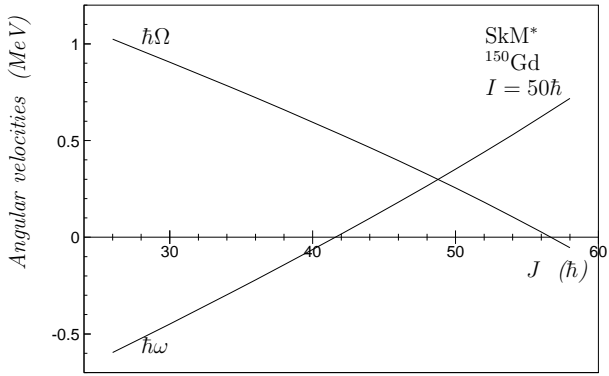}
\end{center}
\caption{Variations of the two relevant angular velocities $\Omega$ and
$\omega$ as functions of the Kelvin circulation $J$ for $I=50\hbar$ solutions.}
\label{fig:4}
\end{figure}

Now to make excursions, for a given value of $I$, out of the yrast solution,
one has to perform a two Lagrange multipliers search, whose result is
examplified in Fig. \ref{fig:4} for the $I=50\hbar$ case. It is rather
significant that to get $J$ values smaller than the yrast value, one should
add a counterotating intrinsic vortical mode. This can be explained by using
the quadratic approximation for the total energy. One finds \cite{Nupha}
\begin{mathletters}
\begin{equation}
I = C \Omega + B \omega
\end{equation}
\begin{equation}
J = B \Omega + A \omega
\end{equation}
\end{mathletters}
where $A$, $B$, and $C$ are (positive) inertia parameters defined in
Ref. \cite{Nupha}. From the above, one finds trivially
\begin{equation}
J = \frac{BI}{C} + \omega \frac{AC-B^2}{C},
\end{equation}
where the coefficient of $\omega$ is found to be positive for well-deformed
nuclei as easily seen from the semiclassical estimates of
Refs. \cite{PRL:stagg,Nupha}. Therefore one can check that starting from
$\Omega>0$ for the yrast state, in order to decrease $J$, one diminishes (from
its vanishing value) the $\omega$ velocity while increasing $\Omega$ (keeping
$I$ constant) so that one gets $\omega\Omega<0$. Conversely the same reasoning
yields $\omega\Omega>0$ when increasing $J$ away from its yrast value.

\begin{figure}
\begin{center}
\epsfig{angle=-90,width=8cm,file=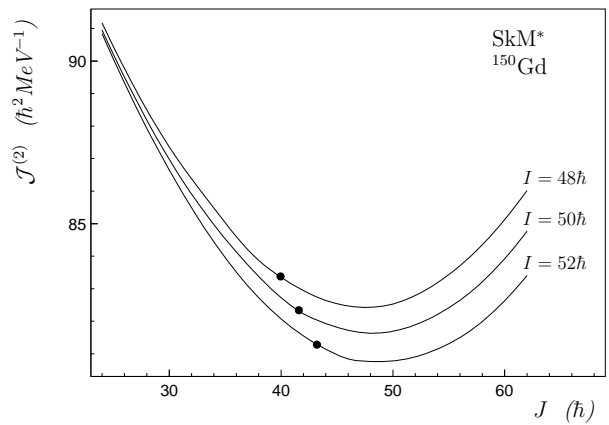}
\end{center}
\caption{Dynamical moment of inertia ${\cal J}^{(2)}$ as a function of the
Kelvin circulation $J$ for angular momentum values around $I=50\hbar$. Dots
correspond to yrast solutions.}
\label{fig:5}
\end{figure}

Pairing correlations act against the global rotations (see, e.g., Ref. 
\cite{Durand}). In our collective model account of such correlations,
this corresponds to a
product of angular velocities such that $\omega\Omega<0$. Consequently
starting from a noncorrelated (Hartree-Fock) solution, the inclusion of
pairing correlations will tend to decrease the Kelvin circulation value (see
Fig. \ref{fig:4}). In
Fig. \ref{fig:5}, the dynamical moments of inertia ${\cal J}^{(2)}$ are
plotted for three values of $I$ around $50\hbar$ as functions of $J$. One sees
that pairing correlations will indeed increase ${\cal J}^{(2)}$ from its
Hartree-Fock value as actually found in the Hartree-Fock-Bogoliubov
calculations of Ref. \cite{Bonche:96}. In this paper, the authors have shown
upon using simple yet realistic pairing matrix elements that the correlations
raise the moment of inertia to the vicinity of the experimental value
(typically ${\cal J}^{(2)} \sim 90 \hbar^2$ MeV$^{-1}$). It is striking that
when constraining the intrinsic vortical mode to obtain this value for ${\cal
J}^{(2)}$, one gets a Kelvin circulation of $\sim 25 \hbar$ which fulfills
precisely the $I/J$ ratio condition of being close to 2. The latter provides a
$4\hbar$ period for the oscillating behavior of the $\gamma$ transition
energies in the $^{150}$Gd region.

To confirm the above conclusion, it would be, of course, very interesting to
perform variational generalized Routhian calculations within the
Hartree-Fock-Bogoliubov approximation. We are currently working on
it. Nevertheless, it seems to us very likely that our indirect estimate already
leads to the conclusion that pairing correlations play the major role in
fine-tuning the staggering period to what is experimentally observed.

\acknowledgments
A part of this work has benefited from a IN2P3 (CNRS)-JINR grant (No. 97-30)
which is gratefully acknowledged.


\begin{thebibliography}{10}

\bibitem{PRL:stagg}
P. Quentin and I.~N. Mikha\"{\i}lov, Phys. Rev. Lett. {\bf 74},  3336  (1995).

\bibitem{Flibotte}
S. Flibotte {\it et~al.}, Phys. Rev. Lett. {\bf 71},  4299  (1993).

\bibitem{Cederwall}
B. Cederwall {\it et~al.}, Phys. Rev. Lett. {\bf 72},  3150  (1994).

\bibitem{Semple}
A.~T. Semple {\it et~al.}, Phys. Rev. Lett. {\bf 76},  3671  (1996).

\bibitem{Haslip}
D.~S. Haslip {\it et~al.}, Phys. Rev. Lett. {\bf 78},  3447  (1997).

\bibitem{Krucken}
R. Kr\"{u}cken {\it et~al.}, Phys. Rev. C {\bf 54},  R2109  (1996).

\bibitem{Hamamoto}
I. Hamamoto and B.~R. Mottelson, Phys. Lett. B {\bf 333},  294  (1994); Phys.
  Scr. {\bf T56}, 27 (1995).

\bibitem{Pavlichenko}
I.~M. Pavlichenkov and S. Flibotte, Phys. Rev. C {\bf 51},  R460  (1995); I.
  M. Pavlichenkov, \textit{ibid.} {\bf 55}, 1275 (1997).

\bibitem{Sun}
Y. Sun, J.-Y. Zhang, and M. Gudrie, Phys. Rev. Lett. {\bf 75},  3398  (1995).

\bibitem{Kota}
V.~K.~B. Kota, Phys. Rev. C {\bf 53},  2550  (1996).

\bibitem{Toki}
H. Toki and L.-A. Wu, Phys. Rev. Lett. {\bf 79},  2006  (1997).

\bibitem{Nupha}
I.~N. Mikha\"{\i}lov, P. Quentin, and D. Samsoen, Nucl. Phys. {\bf A627},  259
  (1997).

\bibitem{Aharonov}
I.~N. Mikha\"{\i}lov and P. Quentin, Eur. Phys. J. A {\bf 1},  229  (1998).

\vskip83ex

\bibitem{Iouldash}
E.~K. Yuldashbaeva, J. Libert, P. Quentin, and M. Girod, Phys. Lett. B (submitted).

\bibitem{Chandra}
S. Chandrasekhar, {\em Ellipsoidal Figures of Equilibrium} (Dover, New York,
  1987).

\bibitem{Rosensteel}
G. Rosensteel, Ann. Phys. (N.Y.) {\bf 186},  230  (1988); Phys. Rev. C {\bf 46},
  1818 (1992).

\bibitem{Routhian}
P. Quentin, D. Samsoen, and I.~N. Mikha\"{\i}lov (unpublished).

\bibitem{Triaxial}
D. Samsoen, P. Quentin, and J. Bartel, Nucl. Phys. {\bf A652} (1999) 34.

\bibitem{Skmstar}
J. Bartel, P. Quentin, M. Brack, C. Guet, and H.-B. Haakansson, Nucl. Phys.
  {\bf A386},  79  (1982).

\bibitem{D1S}
J.-F. Berger, M. Girod, and D. Gogny, Comput. Phys. Commun. {\bf 63},  365  (1991).

\bibitem{Bonche:87}
P. Bonche, H. Flocard, and P.-H. Heenen, Nucl. Phys. {\bf A467},  115  (1987).

\bibitem{Egido}
J.~L. Egido and L.~M. Robledo, Phys. Rev. Lett. {\bf 70},  2876  (1993).

\bibitem{Girod}
M. Girod, J.-P. Delaroche, J.-F. Berger, and J. Libert, Phys. Lett. B {\bf
  325},  1  (1994).

\bibitem{Dudek}
J. Dobaczewski and J. Dudek, Phys. Rev. C {\bf 52},  1827  (1995).

\bibitem{Gall}
B. Gall, P. Bonche, J. Dobaczewski, H. Flocard, and P.-H. Heenen, Z. Phys. A
  {\bf 348},  183  (1994).

\bibitem{Bonche:96}
P. Bonche, H. Flocard, and P.-H. Heenen, Nucl. Phys. {\bf A598},  169  (1996).

\bibitem{Durand}
M. Durand, P. Schuck, and J. Kunz, Nucl. Phys. {\bf A439},  263  (1985).

\end{thebibliography}
\end{document}